\documentclass[12pt]{iopart}

\usepackage{amssymb}
\usepackage{graphicx}

\begin{document}

\title[The r\^ole of correlation entropy in nuclear fusion]{The r\^ole of correlation entropy in nuclear fusion in liquid lithium, indium and mercury}

\author{M. Coraddu}
\address{Istituto Nazionale di Fisica Nucleare (INFN) - Sezione di Cagliari, Monserrato, Italy}
\ead{massimo.coraddu@ca.infn.it}

\author{M. Lissia}
\address{Istituto Nazionale di Fisica Nucleare (INFN) - Sezione di Cagliari, Monserrato, Italy}
\ead{marcello.lissia@ca.infn.it}

\author{P. Quarati}
\address{Istituto Nazionale di Fisica Nucleare (INFN) - Sezione di Cagliari, Monserrato, Italy and
Dipartimento di Scienza Applicata e Tecnologia (DISAT) - Politecnico di Torino, Italy}
\ead{piero.quarati@polito.it}

\author{A.M. Scarfone}
\address{Istituto dei Sistemi Complessi - Consiglio Nazionale delle Ricerche (ISC-CNR) c/o 
Politecnico di Torino, Corso Duca degli Abruzzi 24, I-10129, Italy}
\ead{antoniomaria.scarfone@cnr.it}

\begin{abstract}
Nuclear fusion cross-sections  considerably higher than corresponding theoretical predictions are
observed in low-energy experiments with
metal matrix targets and accelerated deuteron beams.
The cross-section increment is significantly higher for liquid than for solid targets.
We propose that the same two-body correlation entropy used in evaluating the metal melting entropy
explains the large liquid-solid difference
of the effective screening potential that parameterizes the cross-section increment.
This approach is applied to the specific case of the $^6\mathrm{Li}(\mathrm{d},\alpha)^4\mathrm{He}$
reaction, whose measured screening potential
liquid-solid difference is $(235\pm 63)$~eV.
Cross sections in the
two metals with the highest two-body correlation entropy (In and Hg) has not been measured yet:
increments of the cross sections in liquid relative
to the ones in solid metals are estimated with the same procedure.
\end{abstract}

\maketitle

\section{Introduction}

The observed pronounced enhancement of the rate of fusion of deuterons implemented in metal matrices with accelerated deuterons
is well established since the early and almost contemporaneous works by the Kasagi group in
Sendai \cite{Yuki1997,Yuki1998,Kasagi2002}, by the Rolfs group in Gran Sasso and Bochum
\cite{Greife,Raiola2002a,Raiola2002b,Prati:2005gb,Raiola:2005na,Rolfs2006}
and by the Berlin group \cite{Czerski2001,Huke2002,Czerski2008}. Analogous  enhancements has been seen in
fusion between accelerated deuterons
on light nuclei in the target \cite{Fang,Yonemura}.

The understanding of these enhancements of nuclear fusion rates is necessary for the correct evaluation of rates of
stellar fusion processes, of plasma fusion in laboratories and of low-energy cross sections for nuclei in vacuum.
The exact quantitative link between cross sections in different environments (stellar, laboratory, other plasmas, vacuum)
is still under experimental and theoretical scrutiny \cite{Tsyganov2012}.

A standard approach parameterizes these enhancements of the rates adding a constant negative
effective potential to the total energy of the fusing particles.
 Such effective
potentials are much higher than the screening potentials in adiabatic approximation,
which are several tens of eV \cite{Bracci},
while Kasagi et al. \cite{Fang,Yonemura} report
rates of fusion of deuterons on lithium targets
in terms of effective potentials that are several hundreds of eV larger than
adiabatic values.
The same group has been recently measuring rates in a liquid indium matrix, finding a similar behavior \cite{Kasagi2014}.

These potentials can be interpreted as an additional screening, in agreement with the fact that these solid and liquid metals
are dense cold plasmas, where the Debye-H\"uckel model should be modified~\cite{Quarati:2007ie,Quarati:2008hd}.

Several models  appear to reproduce only partially such experimental results.
We have suggested explanations~\cite{Coraddu:2009jw} of the effective potential for fusion in solid metal matrices based on:

1) modifications of the Debye-H\"uckel approach by
spatial fluctuations of the Debye-H\"uckel radius;

2) a critical review of the stopping power;

3)  quantum uncertainty effect which broadens the energy-momentum relation and increases the effective kinetic energy of
the colliding particles following
the Galitzskii-Yakimets quantum description \cite{Galitskii,Eletski,Ferro}.

Furthermore, we have more recently  shown
the importance of the correlation entropy term in metal melting \cite{Quarati2013}.

In the framework where the environment effect is parameterized by adding constant potential to the particles in
the colliding beam, the negative potential is found  \cite{Fang,Yonemura} about
$200-300$~eV  larger in the
  liquid than in the solid metal.
In this paper we show that this difference can be explained by means of the two-body correlation entropy
term of the ions or atoms in liquid phase for the specific case of lithium.
We suggest that this two-body entropy correlation is important and in same case dominant
in fusion processes in
environments different from vacuum. Being the correlation entropy term of In and Hg the
largest among the metals \cite{Wallace1991,Wallace1994}, we evaluate the liquid-solid
difference for these two metals, quantity useful for future experiments.

In Section 2, we give an overview of the experimental situation and outline the standard approaches
used to explain the experimental results.
In Section 3, we report our interpretation of the difference between
the screening potentials in solid and liquid lithium and,
 furthermore, we evaluate the predicted difference for indium and mercury and conclude.

\section{The experimental situation and interpretation of the spread}

In the very-low energy experiments of nuclear fusion in lithium targets, when the center of mass energy
becomes smaller than $30-40$ keV, the cross section is sensibly higher than theoretical vacuum predictions.
The screening potential that parameterizes this enhancement depends on the state of the metal target:

 1) in gaseous targets, one needs screening potentials larger than the electron screening predicted in the adiabatic
limit ($20-30$~eV). In fact the total screening potentials measured by several groups for different gases are in
the range $200-300$~eV~\cite{Fang,Raiola2001};

 2) in the few experiments with liquid target, the anomalous increase of the screening potentials relative to atomic
adiabatic screening plus conduction electron energy in the Thomas-Fermi approximation is even larger.
The total screening potential for lithium is $\approx (520\pm190)$~eV against a prediction of about 71~eV \cite{Fang,Yonemura};

 3) in the many experiments
 with solid targets, effective screening potentials range from 300 to 1300 eV, again values much larger than
the theoretical predictions from atomic and conduction electrons
\cite{Cruz2005,Raiola2006,Huke:2007be,Huke:2008tr,Cruz:2008zz,Cruz2012}.
The specific value
reported for lithium is $\approx (310\pm70)$~eV~\cite{Yonemura}.

4) Kasagi has very recently found \cite{Kasagi2014} anomalous energy spectra of the outgoing $p$ and $t$ in the
 reaction $d\,(d,\,p)\,t$  in liquid Indium.

In summary the atomic electron screening is not enough to explain the enhancement, if its value is
deduced following the standard approach:
adiabatic or sudden approximation and standard Debye-H\"uckel model \cite{Bracci}.
 It appears that target deuterons are not at rest, but they have a finite momentum, and that
reactions are not simple two-body reactions.
Kasagi et al.~\cite{Kasagi2014} suggest that
results could be explained by assuming cooperative colliding
processes; following this hint, we
plan to study possible connections with the quantum momentum-energy uncertainty \cite{Galitskii}.

Attempts to explain the experimental results in Sendai~\cite{Yuki1997,Yuki1998,Kasagi2002}, in
Bochum and Gran Sasso~\cite{Greife,Raiola2002a,Raiola2002b,Prati:2005gb,Raiola:2005na,Rolfs2006},
and in Berlin~\cite{Czerski2001,Czerski2008,Huke2002} are based on:

ionic Debye screening,
assuming liquid lithium is a low-temperature high-density plasma (Toriyabe et al.~\cite{Toriyabe:2012zz});
a simplified model of the
classical quasi-free electrons with an electron screening distance of the order of the 
Debye length, but Debye screening is a cooperative effect of a large number of
electrons inside the Debye sphere and this model needs a number of electrons equal or smaller 
than one (Rolfs et al. \cite{Assenbaum1987,Lattuada2001,Rolfs2006});
use of dynamical screening as from \cite{Shaviv2000} (Dappen et al. \cite{Dappen2012}); 
deuteron dynamics with migration of electrons from the host metallic atoms to the deuterium ones (Huke at al. \cite{Huke:2007be,Huke:2008tr}).

Other authors \cite{Coraddu:2009jw,Eletski,Fisch:2011sb} point out that, as a result of frequent collisions of particles
in environments like metal matrices or dense plasmas, the complete correspondence between total
and kinetic energy disappears and the distribution function depends both on total and kinetic
energy with visible effect on its tail.
Momentum distribution is fundamental for the fusion reaction rates and deviations from
Maxwellian distribution play a central role. Therefore two effects have been examined:
the increase of the deuteron momentum distribution tail due to the quantum energy-momentum
uncertainty effect and the spatial fluctuations of the Debye-H\"uckel radius, the two effects 
lead to larger reaction rates at energies below a few keV.

In liquid indium bombarded with deuteron molecular beam, where anomalous spectra have been
detected, Kasagi et al. \cite{Kasagi2014} introduce a cooperative colliding process.

In the following we shall focus on the difference of screening potential for the $^6$Li(d,$\,\alpha)^4$He
reaction in liquid lithium compared to the same reaction in solid (or atomic/molecular) lithium. Note that
most of the large uncertainty on the two potentials is a
consequence of the large uncertainty on the bare cross section
relative to which the screening enhancement is defined. For instance, the values for liquid and solid
lithium targets from Ref.~\cite{Yonemura} are $520\pm 190$ eV and $310\pm 70$ eV: the difference would
naively be $210\pm 202$ eV. But the uncertainty on the bare cross section cancels in the difference and
the percentage error is much smaller, about $35\%$:
for instance Fang at al.~\cite{Fang} quote $235\pm 65$ eV for the difference between the liquid and
atomic/molecular phase. As a reference typical value we use $200\pm 70$ eV.

\section{Our approach, results and conclusions}
Liquid metals are characterized by a finite microscopic correlation length $L_c$:
spatial correlations or long-range order vanish for distances much greater than $L_c$.
The non-trivial interaction of each particle with the environment is very different from the one of
a quasi-particle in gas: a better description is a particle that moves within a cage
formed by its neighbors; dimensions of this cage are
of order $L_c$ (see Wallace  \cite{Wallace1994,Wallace1989} for a detailed discussion).

We propose that the cooperative contribution of the  $N_{\rm cage}$  atoms in this cage can explain
the additive $\approx 200$~eV effective screening potential in liquid lithium relative to solid lithium
needed
to describe the corresponding increase of the fusion cross section.
The correlation energy of the cage increases the effective energy of the colliding particles: this larger energy can
be equivalently interpreted as a reduction of the stopping power (correlation reduces the transfer of energy to the
environment) or as a direct  contribution of  the cage to the relative kinetic energy of the colliding particles.

In liquid metals the entropy per atom can be written as~\cite{Wallace1989}:
\begin{eqnarray*}
S =S_1+S_2+S_x+S^Q+S^e \ ,
\end{eqnarray*}
where $S_1$ is the one-body (kinetic energy) term, $S_2$ and $S_x$ are the two- and many-body correlation entropies, $S^Q$ is the
 quantum contribution and $S^e$ is due to the electron cloud.
Only the one-body $S_1$  and two-body correlation  $S_2$  entropies
 are large and need to be considered for our scope. In fact, Wallace~\cite{Wallace1991,Wallace1989,Wallace2002}  has calculated  $S_2$
for many liquid metals  as the difference
between the experimental value of the total entropy and the one-body term $S_1$, disregarding corrections from the smaller
terms. In turns, the difference between
the one-body entropy in the two phases is easily determined~\cite{Wallace1991} and it cannot influence
the two-body reaction rate:
only the two-body correlation entropy $S_2$  needs to be considered to explain the different
fusion rates in liquid relative to solid metals.

Our model is the following. When a deuteron enters the metal (solid or liquid) suffers the stopping power process. 
In the solid, at the end of its path, the kinetic energy of the deuteron is reduced by the stopping power process 
and a contribution is gained from the atomic electron screening in the lithium environment. 
The gain can be calculated  by using the modified Debye-H\"uckel approach. 
Furthermore the deuteron collision frequency is increased because of the energy-momentum uncertainty 
(Galitski-Yakimets quantum effect)~\cite{Coraddu:2009jw}. 
Solid lithium has the ions localized in the crystal; the entropy of the crystal is practically the sum of one-body 
contribution $S_1$ (kinetic energy) plus very minor contributions from many-body correlations, 
from  atomic electron cloud and from quantum effects. 
Crystal neither receives or gives energy to the incoming deuteron or to the environment except for the contribution 
given to deuteron from atomic electron cloud (contribution calculated by the adiabatic approximation), 
the ionic screening being negligible (in fact  since the mass of ions is greater than electron mass, 
positive ions have slow mobility and cannot respond quickly to change~\cite{Kasagi2,Toriyabe}). 
When, in the other case, deuteron enters the liquid metal, its interaction with the environment is different for two reasons: 
it travels within a cage of ions/atoms and receives an amount of energy (that we can consider equivalent to a screening energy)
 because the ions in liquid metal, although non localized, are correlated. 
In the liquid metal the negative contribution to entropy of the two-body correlation entropy restores in part a sort of order 
and  the cage can give an amount of energy to the deuteron at the end of the stopping power process or during the travel. 
The internal energy of the cage increases deuteron kinetic energy. 
In summary: deuteron enters the metal sample, travels into the matter submitted to the stopping power process, 
gains energy from the atomic electron (adiabatic approximation and modified Debye-H\"uckel approach), 
increases collision frequency by quantum energy-momentum uncertainty effect and furthermore 
when travels in  the liquid metal collects the amount of energy provided by its cage during or at the end of its itinerary. 
This energy cannot be given to deuteron by the solid metal, is responsible of the different  behavior of Li+d fusion in  solid 
and in liquid metal and represents the difference of the internal energy between liquid and 
solid metal of the cage bunch as quantitatively shown below.

From the energy balance, by calling $\Delta E$  the energy gained by the deuteron and $W$ the energy interaction contributions 
(besides the standard atomic electron screening) due to  interactions of d, Li and Be 
(the d-Li nuclear compound system just before fusion) with the environment, we can set
\begin{eqnarray*}
(W_{\rm d}^{\rm liq} - W_{\rm d}^{\rm sol}) + (W_{\rm Li}^{\rm liq} - W_{\rm Li}^{\rm sol})
   + (W_{\rm Be}^{\rm liq} - W_{\rm Be}^{\rm sol}) -    \Delta E   =  0 \ .
\end{eqnarray*}
While the two differences of the second and third parenthesis are negligible together 
$W_{\rm d}^{\rm sol}$, only  $W_{\rm d}^{\rm liq}$ is non negligible, 
because of the d-cage interaction that assures an amount of energy from the negative two-body correlation entropy term.

Since the free energy is the same in the two phases (solid and liquid), the entropy change must be matched by a change of internal
energy:
\begin{eqnarray}
\Delta U \equiv U^{\rm liq}-U^{\rm sol} =  E_{\rm cage} = \Delta S = N_{\rm cage}\,k_{\rm B}\,T_m\,S_2 \ ,\label{eq:deltaU}
\end{eqnarray}
where we have ascribed  the whole entropy difference to the correlation of the  $N_{\rm cage}$ particles in the cage
and the two-body correlation entropy
$S_2$  is in units of  $k_{\rm B}$. When it is known, it is better to use the
actual temperature of the experiment $T$ instead of
melting temperature $T_m<T$. It would be interesting to test the linear dependence of $ \Delta U$
with the temperature.
Note that the energy
lost locally by the cage is so small that
it does not  modify the macroscopic properties of the liquid.

Therefore $\Delta E \equiv -\Delta U $ is the additional energy available for the colliding particles in liquid
\begin{eqnarray*}
E= E_k+ E_{\rm cage}=E_k+N_{\rm cage}\,k_{\rm B}\,T_m\,S_2 \ ,
\end{eqnarray*}
where $E_k$ is the kinetic energy of the deuteron inside of the metal reduced by the stopping power, whose value can be
calculated following the approach explained in detail by Coraddu et al \cite{Coraddu:2009jw}. Since
the two-body correlation melting entropy is always negative (correlations give order to the system)
in a liquid metal, $E_k=E+\Delta E = E-E_{\rm cage}>E$.

In lithium Wallace \cite{Wallace1991} finds  that $S_2(Li)=-2.48$; given an experimental
temperature $T=520\,K$ and a difference of effective screening potentials
of about $200 \mathrm{\ eV} $,
one finds
\begin{eqnarray}
N_{\rm cage} = \frac{\Delta E}{-S_2 k_BT_m} \approx 1800 \times
\frac{\Delta E}{200\mathrm{\ eV}} \times \frac{2.48}{-S_2} \times \frac{520\,K}{T} \quad .
\label{eq:NcageLi}
\end{eqnarray}

In addition we know a theoretical expression for $S_2$ \cite{Wallace2002}
\begin{eqnarray}
S_2=-{1\over2}\,\rho\int g(r)\,\ln g(r)\,dr\approx-{1\over2}\,\rho\,{V_{cage}\over N_{cage}} \quad ,
\label{eq:S2}
\end{eqnarray}
where $g(r)$ is the two-body correlation function, $V_{\rm cage}$ is the volume occupied by the $N_{\rm cage}$ atoms
of the cage and $\rho$ is the density of the metal in particles per unit volume,
that for liquid lithium is about $4.4\cdot 10^{22}\mathrm{\ cm}^{-3}$. From
Eq.~(\ref{eq:S2}) the
volume of the cage can be estimated
\begin{eqnarray}
  V_{cage} &=& \frac{-2 S_2}{\rho} N_{\rm cage}  =   \frac{2 \Delta E}{\rho k_BT_m} \nonumber \\
&\approx&   2.0 \cdot 10^{-19}  \mathrm{\ cm}^{3}  \times
                   \frac{\Delta E}{200\mathrm{\ eV}} \times  \frac{520\,K}{T}
\quad .
\label{eq:VcageLi}
\end{eqnarray}

Alternatively, if we have a way to calculate/independently measure  the volume of the cage or the
correlation length, we could make predictions and/or tests of the model. For instance,
Ichimaru~\cite{Ichimaru:1993zz} considers liquid metals as strongly coupled plasmas with mean interaction
energy per particle larger than the kinetic energy: the corresponding plasma parameter is much greater
than one. Using Monte Carlo methods Ichimaru determines terms of a power series expansion of the
two-body correlation.

From the theoretical point of view it can be interesting to express the above results in terms of
$x\equiv L_{\mathrm{cage}}/\sigma$, the dimensionless correlation length in units of
the atomic diameter $\sigma$ \cite{Verlet1972}. Expressing the
cage of volume as
$V_{\rm cage}=L_{\rm cage}^3= x^3\sigma^3$ and using Eqs.(\ref{eq:deltaU}), (\ref{eq:NcageLi}),
and (\ref{eq:VcageLi}),
the difference of potentials
\begin{eqnarray}
\Delta U=-{1\over2}\,\rho\,V_{\rm cage}\,k_{\rm B}\,T=-\frac{1}{2} \,\rho\sigma^3\,k_{\rm B}\,T\,x^3 \ ,
\label{eq:deltaUx}
\end{eqnarray}
and the number of atoms in the cage
\begin{eqnarray}
N_{\rm cage}=\frac{1}{2}\frac{\rho\sigma^3}{-S_2}x^3
\label{eq:NcageLix}
\end{eqnarray}
become both functions of the dimensionless parameter $x$.

From this last Eq.~(\ref{eq:NcageLix}) using $\rho\sigma^3\approx 1$, $N_{\rm cage}\approx 1800$ and $-S_2=2.48$, one
finds $x\approx 20$. Under the hypothesis that the entire screening is due to the correlation entropy,
the cage correlation length is about twenty times the diameter of the atom.

If we want to compare and make estimates for other metals,
it is useful to write Eqs. (\ref{eq:deltaU}) and (\ref{eq:deltaUx})
as ratios of the expressions for metal $a$ and for metal $b$:
\begin{eqnarray}
\frac{\left(\Delta U\right)_a}{\left(\Delta U\right)_b} =
                      \frac{\left(N_{\mathrm{cage}}\right)_a}{\left(N_{\mathrm{cage}}\right)_b}
                       \frac{\left(T\right)_a}{\left(T\right)_b}
                       \frac{\left(S_2\right)_a}{\left(S_2\right)_b}
=
                      \frac{\left(\rho\sigma^3\right)_a}{\left(\rho\sigma^3\right)_b}
                       \frac{\left(T\right)_a}{\left(T\right)_b}
                       \frac{\left(x^3\right)_a}{\left(x^3\right)_b}  \quad .
\label{eq:deltaUratio}
\end{eqnarray}

Among the liquid metals, liquid indium and liquid mercury have the highest value
of $S_2$: $S_2(In)=-3.39$ and $S_2(Hg)=-3.97$ with increments of  $+37\%$ and
$+60\%$, respectively, compared to $S_2(Li)=-2.48$. The corresponding
values for $\rho\sigma^3$  are: $\rho\sigma^3(In)=1.37$ and
 $\rho\sigma^3(Hg)=1.12$ with increments of  $+12\%$ and
$-10\%$, respectively, compared to $\rho\sigma^3(Li)=1.25$.

We do not have the experimental values of $\Delta U$ for indium and mercury.
At the same temperature of the lithium experiments $T=520\,K$, assuming that
the dimensionless correlation length $x$ be approximately the same, Eq.~(\ref{eq:deltaUratio})
yields values of $\Delta U$ about $10\%$ larger (smaller) for indium (mercury) relative to the
lithium value. If it
is assumed approximatively equal number of particle in the cage Eq.~(\ref{eq:deltaUratio})
predicts values of  $\Delta U$ much larger, about $40\%$ ($60\%$) larger for indium (mercury)
relative to the lithium value.
In addition Eq.~(\ref{eq:deltaUratio}) predicts that $\Delta U$ is proportional
to the experimental temperature.

In conclusion we have shown that it is possible to explain the difference between the
screening energy to be used in solid and in liquid lithium $\Delta U$ by means of the
 two-body correlation entropy term $S_2$ and we have estimated the possible values of $\Delta U$
for indium and mercury, the two metals with the largest two-body correlation entropy.
In addition
we predict that  $\Delta U$ scales linearly with the temperature.
These evaluations could be useful for future experiments as the ones performed recently  by Kasagi group.


\vspace{5mm}
\noindent
{\bf References}


\begin{thebibliography}{99}
\bibitem{Yuki1997}
H.~Yuki, T.~Satoh, T.~Ohtsuki, T.~Yorita, Y.~Aoki, H.~Yamazaki and J.~Kasagi,
J.\ Phys.\ G:\ Nucl.Part.Phys. {\bf 23}, 1459-1464 ( 1997).

\bibitem{Yuki1998}
H.~Yuki, J.~Kasagi, A.~G.~Lipson, T.~Ohtsuki, T.~Baba,
T.~Noda, B.~F.~Lyakhov, N.~Asami,
JETP lett. {\bf68}, 823-829 (1998).

\bibitem{Kasagi2002}
J.~Kasagi,
H.~Yuki, T.~Baba, T.~Noda, T.~Ohtsuki and A.~G.~Lipson,
J.\ Phys.\ Soc.\ Jpn. {\bf71}, 2281-2285 (2002).

\bibitem{Greife} U. Greife et al., Z. Phys. A {\bf351}, 107 (1995).

\bibitem{Raiola2002a} F. Raiola et al., Eur. Phys. J. A {\bf13}, 377 (2002).

\bibitem{Raiola2002b} F. Raiola et al., Phys. Lett. B {\bf547}, 193 (2002).

\bibitem{Prati:2005gb}
  P.~Prati, F.~Confortola, P.~Corvisiero, H.~Costantini, A.~Lemut, D.~Bemmerer, C.~Broggini and R.~Menegazzo {\it et al.},
  J.\ Phys.\ G {\bf 31}, S1537 (2005).

\bibitem{Raiola:2005na}
  F.~Raiola {\it et al.}  [LUNA Collaboration],
  J.\ Phys.\ G {\bf 31}, 1141 (2005).

\bibitem{Rolfs2006} C. Rolfs, {\em Electron screening in metallic environments: a plasma of the poor man}, Mem. S. A. It. {\bf77}, 907 (2006).

\bibitem{Czerski2001} K. Czerski et al., Europhys. Lett. {\bf54}, 449 (2001).

\bibitem{Huke2002} A. Huke, {\em Die Deuteronen-Fusionsreaktionen in Metallen}, PhD Thesis, (TU Berlin, 2002).

\bibitem{Czerski2008}
  K.~Czerski, A.~Huke, L.~Martin, N.~Targosz, D.~Blauth, A.~G{\'o}rska1, P.~Heide and H.~Winter,
  J.\ Phys.\ G {\bf 35}, 014012 (2008).

\bibitem{Fang} K. Fang et al.,

J. Phys. Soc. Jpn. {\bf80}, 084201 (2011).

\bibitem{Yonemura} H. Yonemura, J. Kasagi, M. Honmo and Y. Isikawa,
{\em Low-energy Li$+d$ reactions with Li target in various phases},
E-print:  http://www.lnl.infn.it/~fusion06/abstract/KASAGI426-kasagi.pdf

\bibitem{Tsyganov2012} E. N. Tsyganov,  Phys.of Atomic Nuclei {\bf 75}, 153 (2012).

\bibitem{Bracci} L. Bracci et al., Nucl. Phys. A {\bf513}, 316 (1990).

\bibitem{Kasagi2014} J. Kasagi, {\em New measurements of screening potential by cooperative colliding        process for the $d+d$ reaction in metallic electron environment}, Research Center for electron photon Science Tohoku University; https://mospace.umsystem.edu/xmlm, (2014).

\bibitem{Quarati:2007ie}
  P.~Quarati and A.~M.~scarfone,
  Astrophys.\ J.\  {\bf 666}, 1303 (2007)
  [arXiv:0705.3545 [astro-ph]].

\bibitem{Quarati:2008hd}
  P.~Quarati and A.~M.~Scarfone,
  J.\ Phys.\ G {\bf 36}, 025203 (2009)
  [arXiv:0811.4053 [nucl-th]].

\bibitem{Coraddu:2009jw}
  M.~Coraddu, M.~Lissia and P.~Quarati,
  Cent.\ Eur.\ J.\ Phys.\ {\bf7}, 527 (2009)
  [arXiv:0905.1618 [nucl-th]].

\bibitem{Galitskii} V. M. Galitskii and V.V. Yakimets, Sov. Phys. JEPT{\bf24}, 637 (1967).

\bibitem{Eletski} A.V. Eletski, A.N. Starostin and M.D. Taran,
 Physics Uspekhi {\bf48}, 281 (2005).

\bibitem{Ferro} F. Ferro and P. Quarati, Phys. Rev. E {\bf71}, 026408 (2008).

\bibitem{Quarati2013} P. Quarati, A.M. Scarfone, Physica A {\bf392}, 6512 (2013).

\bibitem{Wallace1991} D.C. Wallace, Proc. Royal Soc. Lond. A {\bf433}, 631 (1991).

\bibitem{Wallace1994} D.C. Wallace, Int. J. Quant. Chem. {\bf 52}, 425 (1994).

\bibitem{Raiola2001} F. Raiola et al., Eur. Phys. J. A {\bf 10}, 487 (2001).

\bibitem{Cruz2005} J. Cruz et al., Phys. Lett. B {\bf624}, 181 (2005).

\bibitem{Raiola2006} F. Raiola et al., Eur. Phys. J. A {\bf27}, 79 (2006).

\bibitem{Huke:2007be}
  A.~Huke, K.~Czerski and P.~Heide,
  Nucl.\ Instrum.\ Meth.\ B {\bf 256}, 599 (2007)
  [nucl-ex/0701065].

\bibitem{Huke:2008tr}
  A.~Huke, K.~Czerski, S.~M.~Chun, A.~Biller and P.~Heide,
  Eur.\ Phys.\ J.\ A {\bf 35}, 243 (2008)
  [arXiv:0803.1071 [nucl-th]].

\bibitem{Cruz:2008zz}
  J.~Cruz, H.~Luis, M.~Fonseca, Z.~Fulop, G.~Gyurky, F.~Raiola, M.~Aliotta and K.~U.~Kettner {\it et al.},
  J.\ Phys.\ G {\bf 35}, 014004 (2008).

\bibitem{Cruz2012} J. Cruz et al., J.Phys. Conference Series  {\bf 337}, 012062 (2012).

\bibitem{Toriyabe:2012zz}
  Y.~Toriyabe, E.~Yoshida, J.~Kasagi and M.~Fukuhara,
  Phys.\ Rev.\ C {\bf 85}, 054620 (2012).

\bibitem{Assenbaum1987}
 H.~J.~Assenbaum, K.~Langanke and C.~Rolfs,
 Z.\ Phys.\ A\ {\bf 327}, 461-468 (1987).

\bibitem{Lattuada2001}
 M.~Lattuada, R.~G.~Pizzone, S.~Type, P.~Figuera, D. Miljani, A~Musumarra,
 M.~G.~Pellegriti, C.~Rolfs, C.~Spitaleri, and H.~H.~Wolter,
  ApJ {\bf562}, 1076-1080 (2001).

\bibitem{Dappen2012}
  W.~Dappen and K.~Mussack,
  Contr.\ Plasma\ Phys.\ {\bf52}, 149-152 (2012).

\bibitem{Shaviv2000} G. Shaviv and N.J. Shaviv, ApJ {\bf529}, 1054 (2000).

\bibitem{Fisch:2011sb}
  N.~J.~Fisch, M.~G.~Gladush, Y.~.V.~Petrushevich, P.~Quarati and A.~N.~Starostin,
  Eur.\ Phys\. J.\ D {\bf66}, 154 (2012)
  [arXiv:1110.3482 [physics.plasm-ph]].

\bibitem{Wallace1989} D.C. Wallace, Phys. Rev. A {\bf39}, 4843 (1989).

\bibitem{Wallace2002} D.C. Wallace, {\em Statistical Physics of cristals and liquids}, (World Scientific, Singapore, 2002).

\bibitem{Kasagi2} J. Kasagi et al., Nucl. Phys. Rev. {\bf 26 Suppl.}, 44-48 (2009).

\bibitem{Toriyabe} Y. Toriyabe et al., Phys. Rev. C {\bf85}, 054620 (2012).

\bibitem{Ichimaru:1993zz}
  S.~Ichimaru,
  Rev.\ Mod.\ Phys.\  {\bf 65}, 255 (1993).

\bibitem{Verlet1972} L. Verlet, J.J. Weiss, Phys. Rev. A {\bf5}, 939  (1972).

\end{thebibliography}
\end{document}